# Evaluation of research and evolution of science indicators

## Loet Leydesdorff

Université de Lausanne, School of Economics (HEC) & University of Amsterdam, Amsterdam School of Communications Research (ASCoR), Kloveniersburgwal 48, 1012 CX Amsterdam, The Netherlands

**Keywords:** Evaluation, policy decision, research evolution, science indicators.

THE use of scientometric indicators in research evaluation emerged in the 1960s and 1970s, first in the United States and then also in various European countries. Before that time, research evaluation had not been formalized other than through the peer review system, on the one hand, and through economic indicators which could only be used at the macro-level of a national system, on the other. The economic indicators (e.g., percentage of GDP spent on R&D) have internationally been developed by the *Organization of Economic Co-operation and Development* (OECD) in Paris. For example, the *Frascati Manual for the Measurement of Scientific and Technical Activities*[1] can be considered a response to the increased economic importance of science and technology which had become visible in economic statistics during the 1950s.

The idea that scientific knowledge can be organized deliberately and controlled from a mission perspective (for example, for military purposes) was a result of World War II. Before that time the intellectual organization of knowledge had largely been left to the internal mechanisms of discipline formation and specialist communications. The military impact of science and technology through knowledge-based development and mission-oriented research during World War II (e.g. the Manhattan project) made it necessary in 1945 to formulate a new science and technology policy under peacetime conditions.

Vannevar Bush's 1945 report to the US President entitled *The Endless Frontier*[2] contained a plea for a return to a liberal organization of science. Quality control should be left to the internal mechanisms of the scientific elite, for example, through the peer review system. The model of the US National Science Foundation (1947) was followed by other Western countries. [Actually, the first NSF act of 1947 was vetoed by President Truman so that the creation of the NSF was postponed until 1950. The later and more detailed 'Steelman Report', *Science and Public Policy*[3], published in 1947, set in motion the eventual role the federal government would play in supporting fundamental research at universities.] For example, the

e-mail: loet@leydesdorff.net

Netherlands created its foundation for Fundamental Scientific Research (ZWO) in 1950. With hindsight, one can consider this period as the institutional phase of science policies: the main policy instrument was the support of science with institutions to control its funding.

## The *Sputnik* shock

The launching of *Sputnik* by the Soviet Union in 1957 turned the tables for science policy. The Soviets – who had used a non-liberal model – seemed to have become more successful than the West in mission-oriented research. President Eisenhower felt the pressure of the alliance between the scientific elite and the military for enlarging the funding of the science system after the *Sputnik* shock. In addition to his warning (in his farewell speech) against the pressure of a 'military-industrial complex', he formulated as a less known 'second warning':

> Yet in holding scientific research and discovery in respect, as we should, we must also be alert to the equal and opposite danger that public policy could itself become the captive of a scientific-technological elite. [Ref. 4, p. 9]

A far-reaching reorganization of the American research system was one of the consequences. The National Aeronautics and Space Administration (NASA) and the Advanced Research Projects Agency (ARPA), in particular, were created in response to the launch of *Sputnik* and the perception of military threat[5].

During this same period, it became increasingly clear that the continuing growth rates of Western economies could no longer be explained in terms of traditional economic factors such as land, labour, and capital. The 'residue'[6,7] had to be explained in terms of the emerging knowledge-base of the economy[8]. Alongside the military coordination by NATO, the *Organization for Economic Co-operation and Development (OECD)* was created in 1961 in order to organize and to coordinate science and technology policies among its member states, that is, the advanced industrial nations. [The OECD was based on the OEEC, the Organization for European Economic Co-operation, that is, the organization which had served for the distribution of the US and Canadian aid under the Marshall Plan during the postwar period.] As noted, this led in 1963 to the *Frascati Manual* in which parameters were defined for the statistical monitoring of science and technology on a comparative basis. Comparisons among nation states, however, make it possible to raise questions with respect to strengths and weaknesses in the underlying portfolios. During the latter half of the 1960s national S&T policies thus began to emerge in the OECD member states.

For example, the statistics made visible that physics had been successful after World War II in organizing its interests both within the various nation states and at the level of international collaborations (e.g., CERN). Other scientific communities (e.g., molecular biology) claimed more budgetary room given new developments and increases in overall science budgets. During the period 1965–1975, the preferred instrument for dealing with these issues at the national level was a differentiation in the increase rates of budgets at the disciplinary level. In summary, the focus remained on (financial) input-indicators, while the system relied on peer review for more fine-grained decision-making at lower levels[9].

## Output indicators

The attention for the measurement of scientific communication originated from an interest other than research evaluation. During the 1950s and 1960s, the scientific community itself had become increasingly aware of the seemingly uncontrolled expansion of scientific information and literature during the postwar period. In addition to its use in information retrieval, the *Science Citation Index* produced by Eugene Garfield's Institute for Scientific Information came soon to be recognized as a means to objectify standards[10,11]. The gradual introduction of output indicators (e.g., numbers of publications and citations) could be legitimated both at the level of society – because it enables policy makers and science administrators to use arguments of economic efficiency – and internally, because quality control across disciplinary frameworks becomes difficult to legitimate unless objectified standards can be made available in addition to the peer review process.

In 1976 Francis Narin's pioneering study *Evaluative Bibliometrics* was published under the auspices (not incidentally) of the US National Science Foundation[12]. Henry Small had earlier proposed a method for mapping the sciences based on the co-citations of scientific articles[13]. While Small's approach tried to agglomerate specialties into disciplinary structures, Narin focused on hierarchical structures that operate top-down[14,15]. This program appealed to funding agencies like the NSF and NIH that faced difficult decisions in allocating budgets across disciplinary frameworks.

Was it possible to classify the sciences in terms of journal clusters, both substantively and in terms of rank-orders? Might weights be attributed to publications in terms of standards, for example, such as expected versus observed citation rates[16,17]? A scientometric research program at the macro-level could thus increasingly be formulated. While the American studies of the 1970s had focused on the organization of scientific literature, the application of these indicators to research evaluation on an institutional basis was developed in the European context during the 1980s.

Following the publication of Martin and Irvine's study of the relative research performance of various (expensive) installations for radio-astronomy[18], the idea of the assessment of institutional units took hold among policy makers. Leiden University in the Netherlands pioneered during the early 1980s with a fine-grained model for introducing output in terms of publications and recognition through citations as feedback parameters into the finance scheme of departments[19]. This idea was generalized in the UK-model of the Research Assessment Exercises for the funding of university research (since 1992).

The other European countries did not follow the UK in this rationalization of a budget model for research at the national level[20], but pressures prevailed during the 1990s to make publication and citation rates visible in evaluation exercises. For example, after the German unification in 1990, extensive evaluation of the research portfolio of Eastern Germany was immediately placed on the science policy agenda[21]. But can the scientometric indicators carry the political burden of research evaluation?

## Methodological limitations

Publication and citation analyses have become standard tools for research evaluation. However, some methodological problems remain unresolved. The consequent uncertainties have been reflected in hesitations to apply these tools as standards in policy-making processes and research management decisions. How shaky is the ground on which the research evaluations stand?

First, one can raise the question of the unit of analysis in scientific knowledge production and control[22]. The intellectual organization of the sciences does not coincide with their institutional organization. Furthermore, the relations between these layers of organization can be expected to vary among disciplines[23]. The assumption that one can compare 'like with like'[18] in terms of institutional parameters is problematic from the perspective of communication studies[24]. New scientific developments (e.g., artificial intelligence) emerge in very different institutional settings, and in order to make a fair comparison one should perhaps first define a *cognitive* unit of analysis. However, the intellectual organization of the sciences cannot easily be observed or measured[25].

An alternative way to define a unit of analysis would be to base the operationalization on the reflection of scientific developments in the scientific journal literature. The scientific literature is organized in relatively discrete clusters of journals. For example, an article in a biochemistry journal will not often cite an article in condensed matter physics, or vice versa. The relations between these textual units of analysis and the institutional units under evaluation generate further research questions since publication and citation rates differ among disciplines.

The relative decomposability of the literature was central to the above noted attempt of Narin to cluster the database of aggregated journal–journal citations. However, the clustering algorithms provide a snapshot. The structure at any given moment in time does not take into account the dynamic development of the sciences over time. One expects scientific specialties to develop in parallel and not in a hierarchical order. Furthermore, Narin had proposed to *fix* a journal set *ex ante* in order to make comparisons over time possible[12]. However, advanced industrial nations tend to publish in newly emerging areas (and accordingly new journals) relatively more than research units in more conservative systems. The so-called 'decline of British science'– a subject of intense political debate during the 1980s[26] – could with hindsight be deconstructed as partially an artifact of this type of methodological decisions. Within a dynamic database the UK is more stable than in a fixed set, since losses on one side tend to be compensated at the other[27–29].

## Multivariate and dynamic analysis

Journal literature can be considered as a huge network that is knit together in terms of aggregated citations among articles and co-occurrences of words among texts and titles. Factor analysis enables us to hypothesize a latent structure of this network at each moment in time in terms of groupings. However, factor analysis is often computationally too intensive to address the entire database in a single run, let alone to address these questions both comprehensively and dynamically.

Graph analytical techniques are based on recursive ('bottom-up') procedures that operate on relations and can therefore be applied more easily to large datasets. By definition the results of relational analysis, however, can exhibit only relations and hierarchies. The positional analysis of the groupings in a network – that is spanned in terms of relations – is different in kind[30]. But, as noted, the positional analysis remained confined to relatively restricted datasets[31].

A group of researchers at the *École Nationale Supérieure des Mines* in Paris proposed to focus on words and relations among words ('co-words') as an alternative to citation and co-citation analysis[32]. One advantage would be that the words and co-words occur not only in the scientific journal literature, but also in policy reports and patent applications. Can the strengths of the relations among words be used as an indicator of the survival value of an indicated concept during these 'translations' across domains? These authors envisaged that the evaluation of research in terms of performance would become possible by using words and their co-occurrences as indicators of 'translation'[33,34].

The analysis of the co-word patterns proceeded technically in a manner analogous to the co-citation analysis being further developed at the Institute for Scientific Information (ISI) by Small[13]. In the meantime, the ISI group had produced an *Atlas of Science* that was based on agglomerative clustering techniques using a graph analytical algorithm[13,35,36]. These mappings, however, were flawed by the initial decision to focus on hierarchical *relations* for the study of structure and strategic *positions*. Structure can be analysed only in terms of differentiations into latent dimensions of the system. These dimensions can be revealed using factor analytical techniques[37–39].

While the sciences are discursively constructed as networks of communication in terms of relations among words and sentences, the aggregated constructs can be expected to differentiate over longer periods of time according to rules which are functional to the further advancement of the intellectual organization of specialized, and therefore relatively autonomous, structures of scientific communication[24,40,41]. The various discourses continuously update and rewrite reflexively their understandings of the relevant history. This 'self-organization' of the (paradigmatic) discourses takes place from the hindsight perspective of current understanding, both at the level of the active scientists involved and – perhaps with a time lag – at the level of the policy agencies. The latter can direct and influence the development of the sciences only by budgetary and institutional means. The research fronts, however, develop with a communication dynamics different from the institutional organizations.

For example, what one nowadays considers as 'biotechnology' or 'artificial intelligence' is very different from what both policy makers and the scientists involved considered as relevant to these categories in the 1980s when the priority programs in these areas were first developed[42]. The double task of reconstructing the history and of rewriting it as it develops generates the need for a 'double hermeneutics' in the reconstruction[43,44]. Both the observable variation and the selection mechanisms (latent eigenvectors at the network level) can be expected to change over time[45].

Whereas the variation is visible in the data, the selection mechanisms remain latent and can therefore only be hypothesized. On the one hand, these constructs are needed as dimensions for the mapping of the data. On the other hand, constructs remain 'soft', that is, open for debate and reconstruction. De Solla Price's dream of making scientometric mapping a relatively hard social science[46] can with hindsight be considered as fundamentally flawed[47,48]. When both the data and the perspectives are potentially changing, the position of the analyst can no longer be considered as neutral.

During the 1990s, scientometric research evaluation would suffer the kind of fragmentation that is well known to the social scientist. Research evaluation became increasingly contingent on the question of the evaluating agency. The crisis became manifest in the journal *Scientometrics* when this leading journal of the field devoted a special issue[49] to the question of 'Little scientometrics, big scientometrics... and beyond?' in 1994. Not only the reflections became more uncertain than before, but also the subjects under reflection had begun to change because of the increasing focus on the techno-sciences and science-based innovations.

## European Union and the field of research evaluation

Whereas research evaluation was shaped as an agenda at the level of national agencies, the *Single Act* of the European Community in 1986 and the *Maastricht Treaty* of the European Union in 1991 have marked a gradual transition within Europe to a supra-national technology and innovation policy. The EU policies continuously referred to science and technology, because these are considered as the strongholds of the common heritage of the member states. However, the 'subsidiarity' principle prescribes that the European Commission should not intervene in matters that can be left to the nation states. Therefore, a 'federal' research program of the EU could not be developed without taking the detour of a focus on *innovation* as a science-based practice[50].

The *Research, Technology, and Development (RTD) Networks* of the European Union promote transnational and transsectoral collaboration by rewarding the participation of research groups that capitalize on *complementarities* among national origins and institutional spheres. Thus, the operation of 'a triple helix of university–industry–government relations' has been reinforced by the European level of policy making. A system of negotiations and translations among expectations tends to add a dynamic overlay to the nationally institutionalized systems[51].

Interactions can be optimized *ex post* with objectives other than the institutional rationales *ex ante,* and when repeated over time, the network systems increasingly provide their own dynamics. From this perspective, the institutional layer can be considered as the retention mechanism for a network that tends to develop further in terms of its functions. The European Union has provided a feedback on the functions by stabilizing the next-order system's level. The overlay system can be conceptualized as a network mode — named 'Mode 2' by Gibbons *et al.*[52] — or as 'international' when compared with the 'national systems of innovation' previously studied by evolutionary economists[53,54].

The new models can be considered neo-evolutionary insofar as they provide a heuristics for measurement other than the institutional ones that have prevailed at the level of the nation states. For example, national governments have been limitedly successful in developing transdisciplinary programs[55,56]. The bureaucratic focus of Europe, however, legitimated a shift from scientific publications in the traditional format towards achievements and so-called 'deliverables', relatively unhindered by the evaluation schemes of national research councils and scientific communities. These 'deliverables' have become a carrier for next rounds of policy formation in a new mode of research evaluation. From this perspective, the scientific literature can be expected to lag behind the research agenda[57].

Research groups can be sorted by this (trans-national) evaluation scheme in terms of their reliability in providing deliverables to the bureaucracy and therefore in terms of their competencies to serve their audience. In 'Mode 2' research not only the social, but also the intellectual organization of projects and programs is increasingly functionalized in terms of serving relevant audiences[58]. It should be noted that this shift does not imply necessarily a commercialization of science, since the mechanisms remain mainly institutional and therefore non-market[59].

Leydesdorff and Etzkowitz[60] have argued for 'innovation' as the analytical unit of operation of 'innovation systems' that incorporate knowledge-based developments. However, the delineation of a knowledge-based innovation system itself begs the question. Evolutionary economists have emphasized the national character of innovation systems[53,54,61], while a focus on technological developments suggests the sectoral level as the most relevant system of reference[62,63]. Others (e.g., ref. 64) have argued in favour of new technosciences like biotechnology as the frameworks of knowledge integration. The various subsystems of innovation crisscross the organization of society and can be expected to drive (or to inhibit) one another. Thus, the specification of the system of reference for the research evaluation becomes itself increasingly a research question.

## Internet and the development of Cybermetrics and Webometrics

The emergence of the Internet during the 1990s has turned the tables again. Globalization takes place at a supra-institutional level and it reinforces direct relations between science, technology, and the market economy as different mechanisms potentially functional for the coordination. The institutional organizations (e.g., at the national levels) can then be considered as providing niches of communication that develop themselves (or stagnate) by drawing on resources from and by earning credit in the global environments.

Although the carrying organizations provide the original material for their representation at the level of the communication networks, the representations can circulate as 'actants' in the networked relations[65]. The actors behind the actants are increasingly black-boxed when the interactions among the representations begin to resonate at the network level. Thus, the represented systems may become dependent on their representations at the network level, and the carrying systems become reflexively aware of their being reflected[66].

Under these conditions, research evaluation can only position itself reflexively with reference to the representations which are being evaluated because it is analytically unclear (and sometimes not easily accessible) what is precisely reflected and represented[67]. Texts are embedded in *contexts*, but the latter provide the former with meaning given specific codifications of the communication[24]. The different audiences can be served by different discourses using different databases or from different perspectives on the same data. The representations are evaluated in terms of fulfilling their functions at interfaces.

For example, an innovation like the introduction of a new drug on the market has a meaning for the corporation carrying the market introduction which is different from its meaning for the patients suffering from a disease or from that of the scientists and pharmacists who developed the drug. The latter may use *Chemical Abstracts* as their system of reference, or perhaps the *Medline* for searching and reference. The generic name of the drug used in scientific communications will not be familiar to most of its users, who know the same substance only by its tradename. The evaluation schemes of these different audiences can be expected to vary, for example, between the molecular biologists and the medical scientists involved in clinical testing[68].

The agencies carrying these different discourses have only a limited capacity to interface with relevant discourses in their environments ('contexts'). This structuration of the discourses enables each of them to focus on the quality of its own communications. Overarching 'Mode 2' research runs the risk of relating representations to representations without access to the relevant substances because of the formalization. Although the modeling of these complex interactions is a legitimate enterprise, the abstract results of the simulations are very different from the substantive insights of the research evaluations. While the latter can intuitively be understood and translated into policies, the former require first a theoretical interpretation.

## Reflexive scientometrics

Are there still options for evaluating research and other knowledge-based communication given these complex dynamics? When 'All that is solid, melts into air'[69], the melting still leaves behind traces of communication. Communication systems communicate with other communication systems across interfaces and if this mutual information is sufficiently repeated the systems may 'lock-in' and temporarily stabilize into a co-evolution of mutual shaping at an interface. How is one able to operationalize the specificity and thereby the quality of the interfacing communications?

Communication systems (e.g., reflexive discourses) cannot be considered as givens with clear delineations. However, the (sub-)systems of communication can be reconstructed because they have been historically constructed, and insofar their trajectory has been stabilized, this unit of analysis can then also be made amenable to the measurement and eventually management[70]. One is able to observe the events, but they can be provided with meaning only in relations to discourses. However, one can hypothesize these systems of communication on the basis of theoretical information about the specificity of their communication. By raising first the question of 'What is communicated when the system under study communicates?' an analyst is able to reconstruct a code of the communication system under study. Only after addressing this substantive question can one meaningfully proceed to the question of 'how is this codification expected to be communicated?' and then to indicators for the measurement given the specification of the system of reference.

Communications are amenable to measurement because a specific (codified) substance is communicated, and therefore redistributed. A change in the distribution generates a probabilistic entropy (that can be expressed in terms of bits of information). Whereas this mathematical definition of 'information' is still content-free, the specification of a system of reference provides the uncertainty with meaning. As noted, the specification of a system of reference itself is a difficult and analytical task, since one needs information both about the relations of the communications under study with the relevant systems in its environment (at each moment in time) and about the internal development of meaning within the system(s) under study.

Historical research becomes relevant to scientometric analysis from this perspective. The analytical clarification in the reconstructing discourse reproduces the discourses under study as best as it can. Still, the representations may serve purposes other than those of the systems represented. The quality of the representation – both in terms of the analytical specificity and the quantitative precision – becomes crucial to the costs involved in making the management of these representations socially acceptable to the systems represented and thereby to their further development.

Where does this leave us in a seemingly exploding universe of communications and representations? From the perspective of research evaluation, the development of scientometric indicators can only improve the quality of the communication reflexively after asking what a given discourse has been communicating. Qualitative discourses can be expected to be functionally specific. How well has a discourse under study served the internal missions of the field and/or the missions of the agencies who committed the research?

It seems to me that different objectives can be distinguished. First, there is the need at the system's level to provide high-quality information when making decisions about S&T in the public sphere or R&D at institutional levels. This information can only be considered as partial indicators that are locally constructed with reference to specific research questions. However, it is well known that windows of appreciation and evaluation do terrible things to the tangents of the systems under study[71]. One should proceed very carefully in this direction[72].

Second, scientometrics has made us aware that science is amenable to measurement, however imperfect the representation may be. The history of science, the sociology of science, and the philosophy of science can be recognized as qualitative representations of the sciences under study. The relations between qualitative and quantitative approaches can be reformulated: qualitative descriptions and insights inform the measurements as hypotheses and heuristics, while the measurements can be updated and refined by taking interaction terms into account.

Qualitative approaches inform the model from different perspectives, and the results of the quantitative analysis can again be provided with meaning from the various perspectives. The model can then be considered as a machine that enables us further to develop our theories of science. However, as against the modernist concept, the reflexive model can be expected also to fail to carry out this function. All representations remain necessarily incomplete in comparison with the represented system. The sum of the partial representations is not necessarily more informative than their *differences*. Because of this focus on the potential differences in status of the various contributions, the research agenda of quantitative science and technology studies or scientometrics cannot avoid to take a methodological turn: the measurement can no longer avoid the question of what is being measured and why.

## Conclusion

Whereas the research program of the measurement of scientific communications emerged in a context where the delineations among academia, government, and industry were institutionalized, the systemic development of these relations during the second half of the 20th century has changed the system of reference for the evaluation of research. In a knowledge-based economy science fulfills functions that change the definitions of what is consid-ered research and globalization has changed the relevance of national systems of reference. In Europe notably the transnational level has taken the lead in developing innovation policies in an attempt to address the internationalization of industry and the globalization of innovations. Science, of course, has been internationally oriented from its very beginning, but the entrainment of the research process in these global developments is reflected in the research evaluation and the scientometric measurement.

In other words, the systems under study have become more complex. A complex dynamics can analytically be decomposed in several subdynamics. For example, one can raise the question of whether international collaboration in science and coauthorship across national boundaries has emerged during the 1990s as a new subdynamics with characteristics other than domestic collaboration and coauthorship[73,74]. Can a different (e.g., global) subdynamic be hypothesized and then also be measured?[75] Can this new dimension of scientific output also be accounted for in schemes that were developed for institutional management within national systems? The questions generate puzzles at the interfaces between the sciences and the economic and political contexts. The evolving systems and subsystems communicate in different dimensions and the evaluation has become part of the codification of these communications.

1. OECD, *The Measurement of Scientific and Technical Activities: 'Frascati Manual'*, OECD, Paris, 1963.
2. Bush, V., The Endless Frontier: A Report to the President, 1945; Reprinted, Arno Press, New York, 1980.
3. Steelman, J. R., *Science and Public Policy*, Government Printing Office, Washington, DC, 1947. Reprinted by the Arno Press, New York, 1980.
4. York, H. F., *Race to Oblivion: A Participant's View of the Arms Race*, Simon and Schuster, New York, 1970.
5. Edwards, P., *The Closed World, Computers and the Politics of Discourses in Cold War America*, MIT Press, Cambridge, MA, 1996.
6. Abramowitz, M., Resource and output trends in the United States since 1870. *Am. Econ. Rev.*, 1956, **46**, 5–23.
7. OECD, *The Residual Factor and Economic Growth*, OECD, Paris, 1964.
8. Rosenberg, N., *Perspectives on Technology*, Cambridge University Press, Cambridge, 1976.
9. Mulkay, M. J., The mediating role of the scientific elite. *Social Stud. Sci.*, 1976, **6**, 445–470.
10. Price, D. d. S., *Little Science, Big Science*, Columbia University Press, New York, 1963.
11. Elkana, Y., Lederberg, J., Merton, R. K., Thackray, A. and Zuckerman, H., *Toward a Metric of Science: The Advent of Science Indicators*, Wiley, New York, 1978.
12. Narin, F., *Evaluative Bibliometrics: The Use of Publication and Citation Analysis in the Evaluation of Scientific Activity*, National Science Foundation, Washington, DC, 1976.
13. Small, H., Co-citation in the scientific literature: A new measure of the relationship between two documents. *J. Am. Soc. Inf. Sci.*, 1973, **24**, 265–269.
14. Carpenter, M. P. and Narin, F., Clustering of scientific journals. *J. Am. Soc. Inf. Sci.*, 1973, **24**, 425–436.

15. Pinski, G. and Narin, F., Citation influence for journal aggregates of scientific publications: Theory, with application to the literature of physics. *Inf. Proc. Manage.*, 1976, **12**(5), 297–312.

16. Braun, T., Glänzel, W. and Schubert, A., *Scientometric Indicators. A 32-Country Comparative Evaluation of Publishing Performance and Citation Impact*, World Scientific Publications, Singapore/Philadelphia, 1985.

17. Schubert, A. and Braun, T., Standards for citation based assessments. *Scientometrics*, 1993, **26**, 21–35.

18. Martin, B. and Irvine, J., Assessing basic research: Some partial indicators of scientific progress in radio astronomy. *Res. Policy*, 1983, **12**, 61–90.

19. Moed, H. F., Burger, W. J. M., Frankfort, J. G. and van Raan, A. F. J., The use of bibliometric data for the measurement of university research performance. *Res. Policy*, 1985, **14**, 131–149.

20. Hicks, D. and Katz, J. S., Science policy for a highly collaborative science system. *Sci. Public Policy*, 1996, **23**, 39–44.

21. Weingart, P. (ed.), *Die Wissenschaft in osteuropäischen Ländern im Internationalen Vergleich – Eine Quantitative Analyse auf der Grundlage wissenschaftsmetrischer Indikatoren*, Kleine Verlag, Bielefeld, 1991.

22. Collins, H. M., The possibilities of science policy. *Soc. Stud. Sci.*, 1985, **15**, 554–558.

23. Whitley, R. D., *The Intellectual and Social Organization of the Sciences*, Oxford University Press, Oxford, 1984.

24. Leydesdorff, L. and Besselaar, P. v. d., Scientometrics and communication theory: Towards theoretically informed indicators. *Scientometrics*, 1997, **38**, 155–174.

25. Leydesdorff, L., *The Challenge of Scientometrics: The Development, Measurement, and Self-Organization of Scientific Communications*, DSWO Press, Leiden University, Leiden, 1995.

26. Irvine, J., Martin, B., Peacock, T. and Turner, R., Charting the decline of British science. *Nature*, 1985, **316**, 587–590.

27. Leydesdorff, L., Problems with the 'measurement' of national scientific performance. *Sci. Public Policy*, 1988, **15**, 149–152.

28. Braun, T., Gl,,nzel, W. and Schubert, A., Assessing assessments of British science. Some facts and figures to accept or decline. *Scientometrics*, 1989, **15**, 165–170.

29. Martin, B. R., The bibliometric assessment of UK scientific performance – a reply to Braun, Glänzel and Schubert. *Scientometrics*, 1991, **20**, 333–357.

30. Burt, R. S., *Toward a Structural Theory of Action*, Academic Press, New York, 1982.

31. Leydesdorff, L., Can scientific journals be classified in terms of aggregated journal–journal citation relations using the journal citation reports?. *J. Am. Soc. Inf. Sci. Technol.*, forthcoming.

32. Callon, M., Courtial, J.-P., Turner, W. A. and Bauin, S., From translations to problematic networks: An introduction to co-word analysis. *Soc. Sci. Inf.*, 1983, **22**, 191–235.

33. Callon, M., Law, J. and Rip, A. (eds), *Mapping the Dynamics of Science and Technology*, Macmillan, London, 1986.

34. Latour, B., *Science in Action*. Open University Press, Milton Keynes, 1987.

35. Small, H. and Sweeney, E., Clustering the science citation index using co-citations I. A comparison of methods. *Scientometrics*, 1985, **7**, 391–409.

36. Small, H., Sweeney, E. and Greenlee, E., Clustering the science citation index using co-citations II. Mapping science. *Scientometrics*, 1985, **8**, 321–340.

37. Leydesdorff, L., Various methods for the mapping of science. *Scientometrics*, 1987, **11**, 291–320.

38. Leydesdorff, L., A validation study of 'Leximappe'. *Scientometrics*, 1992, **25**, 295–312.

39. Lazarsfeld, P. F. and Henry, N. W., *Latent Structure Analysis*, Houghton Mifflin, New York, 1968.

40. Luhmann, N., *Soziale Systeme. Grundriß einer allgemeinen Theorie*, Suhrkamp, Frankfurt a. M, 1984.

41. Luhmann, N., *Die Wissenschaft der Gesellschaft*, Suhrkamp, Frankfurt a.M, 1990.

42. Nederhof, A. J., Changes in publication patterns of biotechnologists: An evaluation of the impact of government stimulation programs in six industrial sectors. *Scientometrics*, 1988, **14**, 475–485.

43. Giddens, A., *New Rules of Sociological Method*, Hutchinson, London, 1976.

44. Giddens, A., *The Constitution of Society*, Polity Press, Cambridge, 1984.

45. Leydesdorff, L., Why words and co-words cannot map the development of the sciences. *J. Am. Soc. Inf. Sci.*, 1997, **48**(5), 418–427.

46. Price, D. de Solla, Editorial statement. *Scientometrics*, 1978, **1**(1), 7–8.

47. Wouters, P. and Leydesdorff, L., Has Price's dream come true: Is scientometrics a hard science?. *Scientometrics*, 1994, **31**, 193–222.

48. Price, D. d. S., Citation measures of hard science, soft science, technology, and nonscience. In *Communication among Scientists and Engineers* (eds Nelson, C. E. and Pollock, D. K.), Lexington, MA, Heath, 1970, pp. 3–22.

49. Glänzel, W. and Schöpflin, U., Little scientometrics, big scientometrics.... and beyond. *Scientometrics*, 1994, **30**(2–3), 375–384.

50. Narin, F. and Noma, E., Is technology becoming science?. *Scientometrics*, 1985, **7**, 369–381.

51. Etzkowitz, H. and Leydesdorff, L., The dynamics of innovation: From national systems and 'mode 2' to a triple helix of University–Industry–Government relations. *Res. Policy*, 2000, **29**(2), 109–123.

52. Gibbons, M., Limoges, C., Nowotny, H., Schwartzman, S., Scott, P. and Trow, M., *The New Production of Knowledge: The Dynamics of Science and Research in Contemporary Societies*, Sage, London, 1994.

53. Lundvall, B.-Å. (ed.), *National Systems of Innovation*, Pinter, London, 1992.

54. Nelson, R. R. (ed.), *National Innovation Systems: A Comparative Analysis*, Oxford University Press, New York, 1993.

55. Van den Daele, W., Krohn, W. and Weingart, P. (eds), *Geplante Forschung: Vergleichende Studien über den Einfluss politischer Programme auf die Wissenschaftsentwicklung*, Suhrkamp, Frankfurt a.M., 1979.

56. Studer, K. E. and Chubin, D. E., *The Cancer Mission. Social Contexts of Biomedical Research*, Sage, Beverly Hills, 1980.

57. Lewison, G. and Cunningham, P., Bibliometric studies for the evaluation of trans-national research. *Scientometrics*, 1991, **21**, 223–244.

58. Kobayashi, S.-I., Applying audition systems from the performing arts to R&D funding mechanisms: Quality control in collaboration among the academic, public, and private sectors in Japan. *Res. Policy*, 2000, **29**(2), 181–192.

59. Nowotny, H., Scott, P. and Gibbons, M., *Re-Thinking Science: Knowledge and the Public in an Age of Uncertainty*, Polity, Cambridge, 2001.

60. Leydesdorff, L. and Etzkowitz, H., The triple helix as a model for innovation studies. *Sci. Public Policy*, 1998, **25**(3), 195–203.

61. Skolnikoff, E. B., *The Elusive Transformation: Science, Technology and the Evolution of International Politics*, Princeton University Press, Princeton, NJ, 1993.

62. Pavitt, K., Sectoral patterns of technical change: Towards a theory and a taxonomy. *Res. Policy*, 1984, **13**, 343–373.

63. Freeman, C., Continental, national and sub-national innovation systems – Complementarity and economic growth. *Res. Policy*, 2002, **31**, 191–211.

64. Carlsson, B. (ed.), *New Technological Systems in the Bio Industries – An International Study*, Kluwer Academic Publishers, Boston/Dordrecht/London, 2002.

65. Callon, M. and Latour, B., Unscrewing the big leviathan: How actors macro-structure reality and how sociologists help them to do

so. In *Advances in Social Theory and Methodology. Toward an Integration of Micro- and Macro-Sociologies* (eds Knorr-Cetina, K. D. and Cicourel, A. V.), Routledge & Kegan Paul, London, 1981, pp. 277–303.

66. Wouters, P., The citation culture. Unpublished Ph D thesis, University of Amsterdam, 1999.

67. Rip, A., Qualitative conditions for scientometrics: The new challenges. *Scientometrics*, 1997, **38**(1), 7–26.

68. Leydesdorff, L., Indicators of innovation in a knowledge-based economy. *Cybermetrics*, 2001, **5**(1), Paper 2, at http://www.cindoc.csic.es/cybermetrics/articles/v5i1p2.html (Accessed on 30 July 2005).

69. Marx, K., *The Communist Manifesto*, Paris, Translated by Samuel Moore in 1888. Penguin, Harmondsworth, 1967, 1848.

70. Leydesdorff, L., *A Sociological Theory of Communication: The Self-Organization of the Knowledge-Based Society*, Universal Publishers, Parkland, FL, 2001; at < http://www.upublish.com/books/leydesdorff.htm >; (Accessed on 30 July 2005).

71. Casti, J., *Alternate Realities*. Wiley, New York, 1989.

72. Van Raan, A. F. J. (ed.), *Handbook of Quantitative Studies of Science and Technology*, Elsevier, Amsterdam, 1988.

73. Persson, O., Glänzel, W. and Danell, R., Inflationary bibliometric values: The role of scientific collaboration and the need for relative indicators in evaluative studies. *Scientometrics*, 2004, **60**(3), 421–432.

74. Wagner, C. S. and Leydesdorff, L., Mapping the network of global science: Comparing international co-authorships from 1990 to 2000. *Int. J. Technol. Globalization*, 2005, **1**(2), 185–208.

75. Wagner, C. S. and Leydesdorff, L., (forthcoming). Network structure, self-organization and the growth of international collaboration in science. *Res. Policy*, 2005 (in print).

---

## Quotations on Eugene Garfield

'...one of the remarkable things about Eugene Garfield is that along with the imagination, pragmatic judgment, and immense energy required to invent, produce, and develop a useful tool for a superficially routine, but fundamental task in science – searching the literature – he has a deep intuitive sense of the social, cultural, and cognitive structures latent in the practice of science'.

— Blaise Cronin and Helen Barsky Atkins
In *The Web of Knowledge – A Festschrift in Honour of Eugene Garfield*
Information Today Inc, Medford, 2000

'Arguably, Eugene Garfield's lasting legacy will not be the pioneering products he conceived and launched, notably the Science Citation Index (SCI) and its siblings, the ever popular Current Contents, or, most recently, the newspaper, The Scientist. Nor will it be the highly successful company he founded in Philadelphia in 1958, which still trades as the Institute for Scientific Information (ISI). Rather it will be the inter-disciplinary research community, denoted by the portmanteau label "scientometrics", which has evolved and cohered over the course of the last four decades....'

— Blaise Cronin and Helen Barsky Atkins
In *The Web of Knowledge – A Festschrift in Honour of Eugene Garfield*
Information Today Inc, Medford, 2000

'Without the drive, perseverance and social capacities of Eugene Garfield as well as the technical expertise with which he surrounded himself, the immense task of building the SCI would probably not even have been thinkable'.

— Wouters, P.
*The Citation Culture*
Ph D Dissertation, University of Amsterdam, Amsterdam, 1999

'Some thirty-eight years later, under the rubric of NIH/Harold Varmus' "Pubmed Central" we can foresee the exploitation of modern computer and communication technology to the efficient service of the needs of science and of the public. No one has toiled with greater diligence and insight to these ends than Eugene Garfield.'

— Joshua Lederberg
In *The Web of Knowledge – A Festschrift in Honour of Eugene Garfield*
Information Today Inc, Medford, 2000